# Two-dimensional antiferroelectric tunnel junction

Jun Ding,[1,2,*] Ding-Fu Shao,[1,*,†] Ming Li,[1] Li-Wei Wen,[2] and Evgeny Y. Tsymbal[1,‡]

[1] *Department of Physics and Astronomy & Nebraska Center for Materials and Nanoscience, University of Nebraska, Lincoln, Nebraska 68588-0299, USA*

[2] *College of Science, Henan University of Engineering, Zhengzhou 451191, People's Republic of China*

Ferroelectric tunnel junctions (FTJs), which consist of two metal electrodes separated by a thin ferroelectric barrier, have recently aroused significant interest for technological applications as nanoscale resistive switching devices. So far, most of existing FTJs have been based on perovskite-oxide barrier layers. The recent discovery of the two-dimensional (2D) van der Waals ferroelectric materials opens a new route to realize tunnel junctions with new functionalities and nm-scale dimensions. Due to the weak coupling between the atomic layers in these materials, the relative dipole alignment between them can be controlled by applied voltage. This allows transitions between ferroelectric and antiferroelectric orderings, resulting in significant changes of the electronic structure. Here, we propose to realize 2D antiferroelectric tunnel junctions (AFTJs), which exploit this new functionality, based on bilayer $In_2X_3$ (X = S, Se, Te) barriers and different 2D electrodes. Using first-principles density functional theory calculations, we demonstrate that the $In_2X_3$ bilayers exhibit stable ferroelectric and antiferroelectric states separated by sizable energy barriers, thus supporting a non-volatile switching between these states. Using quantum-mechanical modeling of the electronic transport, we explore in-plane and out-of-plane tunneling across the $In_2S_3$ van der Waals bilayers, and predict giant tunneling electroresistance (TER) effects and multiple non-volatile resistance states driven by ferroelectric-antiferroelectric order transitions. Our proposal opens a new route to realize nanoscale memory devices with ultrahigh storage density using 2D AFTJs.

Electron tunneling is a quantum-mechanical phenomenon where electrons are transmitted across a potential barrier exceeding their energy. The investigation of this phenomenon in material science has offered a route toward useful electronic devices, such as tunnel junctions which consist of two metallic electrodes separated by a thin insulating barrier layer [1]. Ferroelectric (FE) insulators are promising barrier materials, due to their spontaneous electric polarization, which can be switched between two orientations by an external electric field, resulting in the tunneling electroresistance (TER) effect [2,3]. The TER effect manifests itself in a large resistance change with polarization reversal and thus is interesting for potential applications of ferroelectric tunnel junctions (FTJs) in non-volatile information storage and processing [4-8].

The most common origin of TER is incomplete screening of the polarization charge at barrier/electrode interfaces [4,7,8]. This produces a depolarizing field, affecting the electrostatic potential profile in a FTJ. The asymmetry in the potential profile and hence in the effective barrier height for different FE polarization orientations leads to the TER effect (Fig. 1a). So far, studies of FTJs have been focused on engineering the electrode and interface materials in order to have a larger change in the effective barrier height with polarization switching to enhance the TER effect. It has been demonstrated that a sizable TER effect can be achieved by using dissimilar electrodes [9- 17] or by controlling FTJ interfaces [18-21].

The choice of suitable electrode materials for FTJs is however often problematic. On one hand, large TER requires significant difference in the electrode properties (such as their screening lengths). On the other hand, different chemical potentials of the two electrodes produce a strong built-in electric field across the FTJ, which often prevents polarization switching. A more active role of the barrier can be achieved by realizing non-uniform polarization states [22-25]. However, this requires specific experimental set-up and large barrier thickness. A direct control of the tunneling barrier by tuning the band structure of the FE insulator, would be a more efficient way to enhance the performance of FTJs. The TER effect in such a FTJ with the ferroelectrically tunable electronic structure in the barrier would not rely on the electrode materials, but rather on the FE barrier itself. However, the conventional FE materials, where different polarization states are topologically identical, exhibit the same electronic structure.

The band structure change in an insulator could be realized by the control of a long-range electric dipole order. Specifically, switching between FE and antiferroelectric (AFE) phases [26], i.e. a transition between parallel and antiparallel orientations of the electric dipoles, is expected to change the electronic band structure of the material. However, this type of transition is usually induced by temperature or by a strong electric field [27,28]. In the former case, the transition is not isothermal, as required for device applications, whereas in the latter case, the FE-like state is volatile, i.e. can be stabilized only under the influence of the applied electric bias. For example, AFE tunnel junctions (AFTJ) based on AFE $PbZrO_3$ barrier layers showed a very large TER effect (up to $10^9$ % at room temperature) associated with the transition between non-polar AFE and polar FE states under applied bias voltage [29]. However, the polar state was sustained only in the presence of applied electric field. In fact, a non-volatile AFE-FE switching has never been realized in the conventional three-dimensional (3D) ferroic insulators, due to the strong bonding across the adjacent layers.

The recent discovery of the two-dimensional (2D) van der Waals FE materials opens a route to realize this property

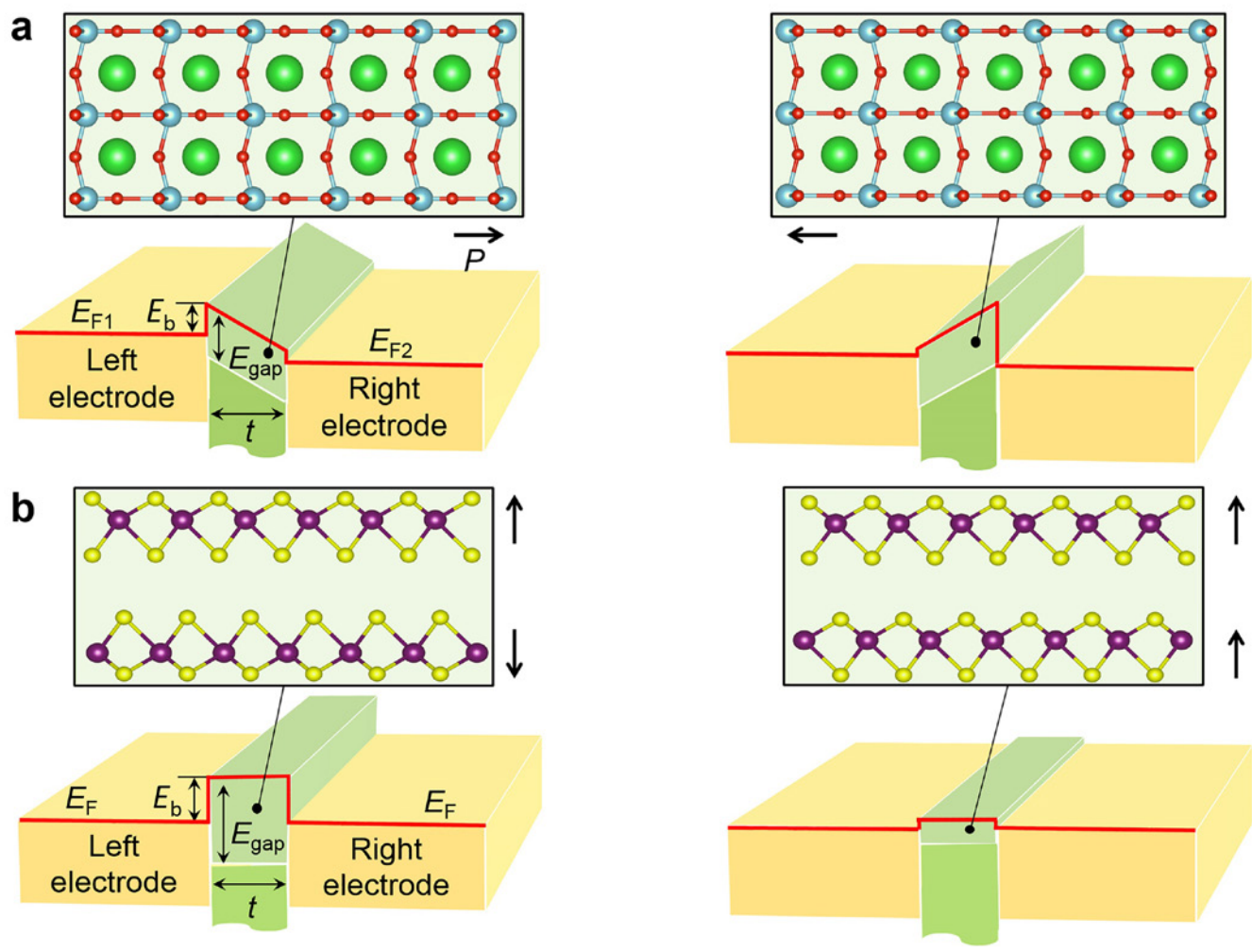

**FIG. 1 (a)** A conventional FTJ with a 3D FE barrier material. Two different electrodes are used to produce asymmetry in the electrostatic potential profile when the FE polarization is reserved. **(b)** An AFTJ with a 2D bilayer barrier layer made of a van der Waals FE insulator. The barrier height changes due to an FE-AFE switching.

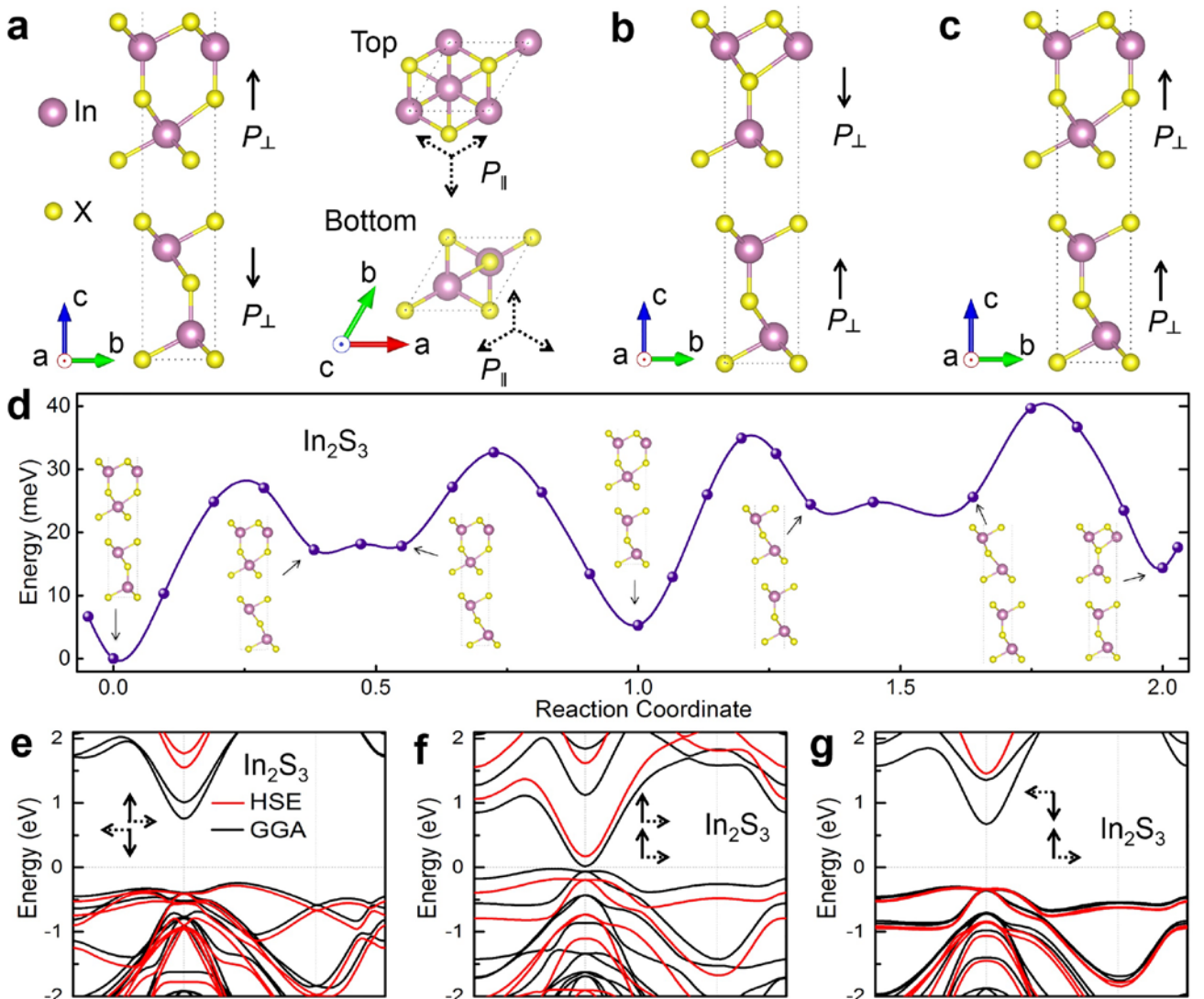

**FIG. 2. (a, b, c)** Crystal structure of bilayer $In_2X_3$ (X = S, Se, Te) in the AFE-T **(a),** AFE-H **(b)** and FE **(c)** states. The right panel of (a) shows the top views of each layer of the AFE-T state. Solid arrows denote the out-of-plane polarization. Dashed arrows denote the in-plane polarization along the three identical in-plane polar directions. **(d)** Total energy along the transition path between the three polar states of bilayer $In_2S_3$. The atomic displacement between different polarization states is parameterized by a reaction coordinate, where 0, 1, and 2 represent the AFE-T, FE, and AFE-H states, respectively. **(e, f, g)** The calculated band structures of $In_2S_3$ in the AFE-T (e), AFE-H (f) and FE states (g) using GGA (black lines) and HSE (red lines).

[30,31]. Ferroelectricity has been experimentally demonstrated in $CuInP_2S_6$ [32,33], SnTe [34-36], and $In_2Se_3$ [37- 43]. These 2D materials have layered structures, where the interlayer coupling is much weaker than that in the conventional 3D materials. In a few-layer 2D ferroelectric, the electric dipole alignment is maintained nearly independently within each layer, so that parallel and antiparallel dipole orientations between the layers have similar energies. This property could support non-volatile switching between the AFE and FE states. In this case, the FE state would produce a depolarizing field resulting in a relative band energy shift across the barrier layer, which is absent in the AFE state. These changes in the electronic structure would inevitably affect the transport properties of the AFTJ through, e.g., changing the barrier height (Fig. 1b).

From the practical perspective, $In_2Se_3$ is especially interesting. This 2D FE insulator hosts intrinsically intercorrelated out-of-plane and in-plane polarization [30,38] and exhibits coexisting FE and AFE domains [38,44], as has been experimentally observed in trilayer structures [37]. These results indicate a possibility of using $In_2Se_3$ as a barrier layer in a 2D AFTJ, where the AFE-FE phase transition can be achieved to directly control the tunneling barrier height.

In this work, we exploit these properties of $In_2Se_3$ and related chalcogenides to realize a 2D AFTJ which functional properties are controlled by the AFE-FE phase transitions. Using first-principles density functional theory (DFT) calculations [47], we demonstrate that the $In_2X_3$ (X = S, Se, Te) bilayers exhibit stable FE and AFE states separated by sizable energy barriers, thus supporting a non-volatile switching between these states. We further explore in-plane and out-of-plane tunneling across the $In_2S_3$ bilayers in AFTJs with different electrodes, and predict giant TER effects and multiple non-volatile resistance states driven by FE-AFE order transitions.

$In_2Se_3$ has several different structural phases [62,63]. The FE phase observed in $In_2Se_3$ belongs to the $R3m$ space group, which consists of the rhombohedral stacking of $In_2Se_3$ layers [30,37]. Each $In_2Se_3$ layer contains five triangular lattices stacked with Se-In-Se-In-Se sequence, as shown in Figs. 2(a-c). The atoms within each layer are connected by covalent bonds, while the different layers are coupled by the van der Waals interaction. Two topologically identical polar states can be switched by a locked out-of-plane and in-plane motion of the middle Se atom. This polar displacement produces a finite out-of-plane polarization $P_\perp$ along the $z$ direction, and three equivalent in-plane polarizations $P_\parallel$ along the $[110]$, $[\bar{2}10]$, and $[1\bar{2}0]$ directions due to the three-fold rotation symmetry of the $R3m$ space group. The sum of the three in-plane polarizations leads to a zero in-plane net polarization. However, device geometry or a substrate proximity effect might break the three-fold rotation symmetry, leading to a net in-plane polarization as observed in experiments [38,40].

Although the FE structure has not yet been detected in bulk $In_2S_3$ and $In_2Te_3$, the phonon calculations indicated that the $In_2X_3$ (X = S, Se, Te) family is stable in this structural phase in the 2D limit [30]. As shown in Figs. 2(a-c), there are three different types of the dipole ordering in bilayer $In_2X_3$, i.e. a tail-to-tail AFE state (denoted as AFE-T), a head-to-head AFE state (denoted as AFE-H), and a FE state. The only difference between these states is the position of the middle X atoms in the bottom and top layers.

To explore the electronic properties and switching between these states, we use $In_2S_3$ as a representative material. We consider a transition path between these states as shown in Fig. 2(d), where the AFE-T state transforms to the FE state through two metastable states associated with the atomic displacement in the bottom layer, as suggested in Ref. 30. Similarly, the FE state transforms to the AFE-H state through two metastable states associated with the atomic displacement in the top layer. The transition barrier between these states of about 30-40 meV is comparable to the FE switching barriers of typical 3D perovskite oxides [64]. All the three states appear to be energy minima. Consistent with the previous results [30,37,38], we find that the AFE-T state has the lowest energy and the AFE-H state has the highest energy. This is due the increasing negative charges of the interfacial S anions in the AFE-H structure and hence the enhancement of the Coulomb repulsion [47]. The calculations for $In_2Se_3$ and $In_2Te_3$ demonstrate similar results [47]. The energy difference between these states could be reduced by building an appropriate heterostructure [65]. It is notable that in all $In_2X_3$ bilayers, the FE polarization is about 1 μC/cm$^2$, which is much smaller than that of most FE perovskite oxides (~10-100 μC/cm$^2$) [47].

Figs. 2(e-g) (black lines) show the band structures of the $In_2S_3$ bilayer in the AFE and FE states calculated using generalized gradient approximation (GGA) [66]. The band structure reveals an indirect band gap, where the conduction band minimum (CBM) is located at the Γ point, whereas the valence band maximum (VBM) is located in the Γ-K direction for the AFE-T state and in the Γ-M direction for the AFE-H and FE states. The two AFE states have moderate band gaps of about 1 eV, while the FE state has a very small band gap of 0.037 eV. The latter is due to the out-of-plane polarization in the FE state, which produces a depolarizing electric field across the bilayer, resulting in the relative shift of the energy bands of the two $In_2S_3$ layers. The large difference between the AFE and FE band gaps is further confirmed by electronic structure calculations using the Hyed-Scuseria-Ernzerhof (HSE) hybrid exchange-correlation functional (red lines in Figs. 2(e-g)), which is considered to be more accurate for the energy band gap prediction [67]. The possibility of AFE-FE switching with the associated substantial changes in the electronic band structure makes an $In_2X_3$ bilayer a promising barrier material to realize a 2D AFTJ.

First, we consider a *symmetric in-plane* AFTJ, where an $In_2S_3$ bilayer barrier separates identical Cd-doped $In_2S_3$ bilayer electrodes, $(Cd_{0.5}In_{0.5})_2S_3$, and electron transport occurs parallel to the plane of the junction. Since Cd has one valence electron less than In, $(Cd_{0.5}In_{0.5})_2S_3$ can be considered as the *p*-doped $In_2S_3$, where the Fermi level ($E_F$) is shifted below the VBM by hole doping. The tunneling barrier of about 7 nm in width is constructed by stacking 10 orthorhombic unit cells of $In_2S_3$ along the [110] direction, as shown in Fig. 3 and Fig. S4 [47]. We fix the electrode to the AFE-T structure. When the barrier is in the AFE-T state, the large band gap is well preserved across the whole barrier, as seen from the layer resolved density of states (LDOS) in Fig. 3(b). When the barrier is in the AFE-H state, there are metallic states at the electrode/barrier interfaces (Fig. 3(c)), due to domain walls (Fig. S4). This does not affect the large band gap in the center of the barrier (Fig. 3(c)). On the contrary, when the barrier is in the FE state, the band gap is strongly reduced, as is evident from Fig. 3(d). In this case, the band edges, i.e. the CBM and the VBM, vary monotonically across the FE $In_2S_3$ barrier. This is due to the AFTJ geometry breaking the three-fold rotation of $In_2S_3$ and leading to a finite net polarization along the [110] direction. This net in-plane polarization produces polarization charges of opposite sign at the two interfaces between the barrier and the electrodes, resulting in the depolarizing electric field and the associated band bending across the barrier region. Due to a small band gap of the FE $In_2S_3$, the band bending causes the $E_F$ to cross the VBM and CBM of $In_2S_3$ near the left and right interfaces, respectively (Fig. 3(d)).

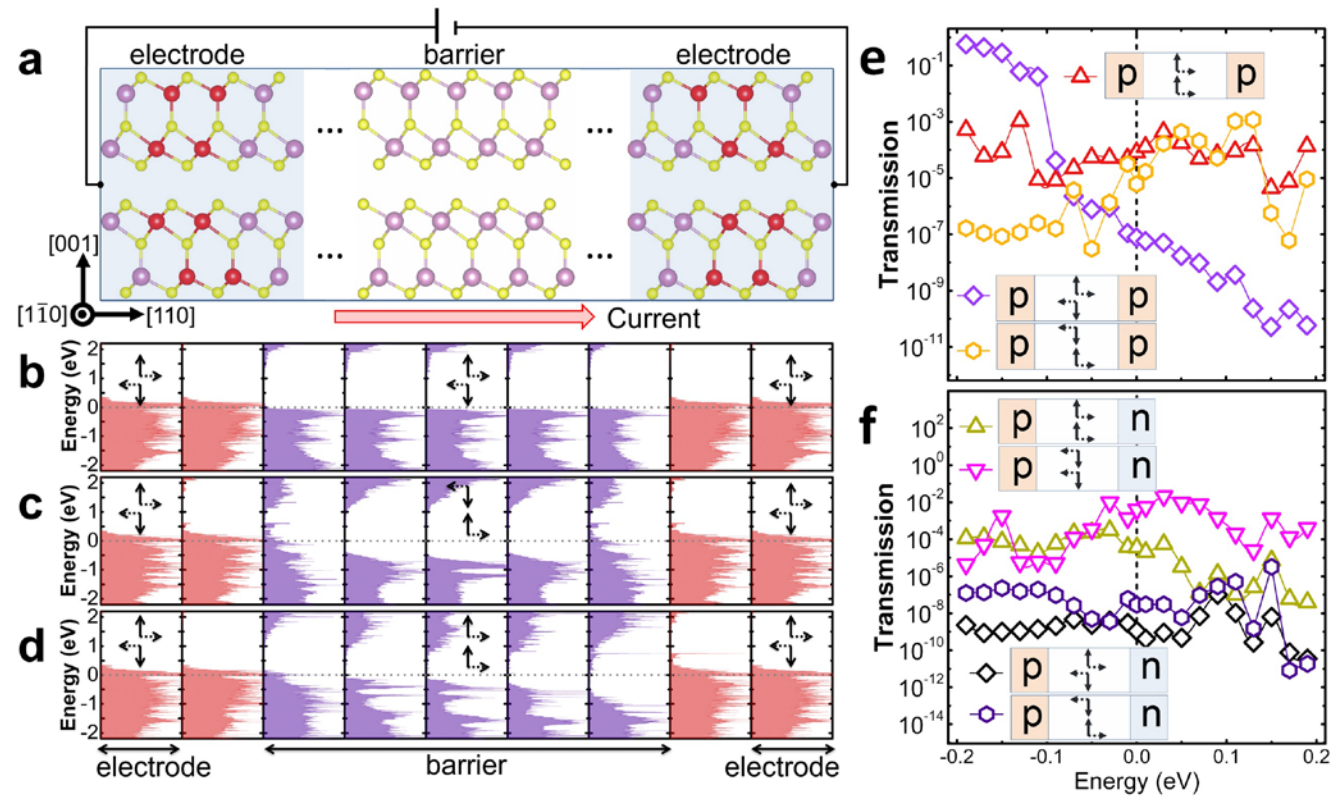

**FIG. 3.** **(a)** Atomic structure of the $(Cd_{0.5}In_{0.5})_2S_3/In_2S_3/(Cd_{0.5}In_{0.5})_2S_3$ in-plane AFTJ in the AFE-T state. **(b-d)** The layer resolved density of states of the $(Cd_{0.5}In_{0.5})_2S_3/In_2S_3/(Cd_{0.5}In_{0.5})_2S_3$ AFTJ in the AFE-T state (b), AFE-H state (c), and FE state (d). The layers of panels (b)-(d) are shown in Fig. S4. **(e,f)** Transmission as a function of electron energy for the $(Cd_{0.5}In_{0.5})_2S_3/In_2S_3/(Cd_{0.5}In_{0.5})_2S_3$ **(e)** and $(Cd_{0.5}In_{0.5})_2S_3/In_2S_3/(Sn_{0.5}In_{0.5})_2S_3$ **(f)** in-plane AFTJs.

These features of the electronic band structure of the in-plane $(Cd_{0.5}In_{0.5})_2S_3/In_2S_3/(Cd_{0.5}In_{0.5})_2S_3$ AFTJ are reflected in the calculated transmission. When the AFTJ is in the AFE-T state, transmission $T_{\text{AFE-T}}$ is exponentially reduced with increasing energy due to decreasing proximity of the VBM (Fig. 3(e)). When the AFTJ is in the AFE-H state, since $E_F$ is deep inside the band gap, transmission $T_{\text{AFE-H}}$ is gradually enhanced with increasing energy due to the decreasing effective barrier height (Fig. 3(e)). On the contrary, when the AFTJ is in the FE state, the transmission $T_{\text{FE}}$ is weakly dependent on energy (Fig. 3(e)). This is due to the band bending across $In_2S_3$, which causes the effective barrier height to be nearly independent of energy. We find that although the metallic domain walls reduce the effective barrier width for the AFTJ in the AFE-H state and thus enhance $T_{\text{AFE-H}}$ compared to $T_{\text{AFE-T}}$, the $T_{\text{AFE-H}}$ is still smaller than $T_{\text{FE}}$ at energies near $E_F$, due to the enhanced barrier height for the former. At $E = E_F$, the predicted on/off ratios for the AFTJ are as large as $T_{\text{FE}}/T_{\text{AFE-T}} \sim 10^4$ and $T_{\text{AFE-H}}/T_{\text{AFE-T}} \sim 10^2$. For $E = E_F + 0.2$eV, the $T_{\text{FE}}/T_{\text{AFE-T}}$ is enhanced up to $\sim 10^7$, which can be achieved by appropriate engineering of the band alignment between the electrodes and the insulating barrier.

For the symmetric AFTJs considered so far, $T_{\text{FE}}$ is the same for polarization of $In_2S_3$ pointing up or down due to identical electrodes. Therefore, such symmetric AFTJs exhibit three non-equivalent resistance states, corresponding to the FE, AFE-T, and AFE-H states of the barrier. Using different electrodes is expected to reveal properties of a conventional FTJ where screening of the non-vanishing in-plane polarization by asymmetric electrodes results in different transmission depending on polarization orientation (such as in 2D tunnel junctions studied in Refs. [68-70]). Combination of the two mechanisms in the AFTJ, i.e. change in the band structure by the dipole ordering and modulation of the barrier height by the asymmetric screening, is expected to result in four non-equivalent resistance states. To demonstrate this property, we consider an *asymmetric* in-plane AFTJ where the two bilayer electrodes are different: $(Cd_{0.5}In_{0.5})_2S_3$ and $(Sn_{0.5}In_{0.5})_2S_3$, where the latter can be considered as *n*-doped $In_2S_3$. As expected, the proposed asymmetric AFTJ allows realizing a switch with four non-volatile resistance states, each of them being distinguished by transmission different by several orders in magnitude, i.e. $T_{\text{AFE-H}}/T_{\text{AFE-T}} = 26.1$, $T_{\text{FE-up}}/T_{\text{AFE-T}} = 3.2\times10^4$, and $T_{\text{FE-down}}/T_{\text{AFE-T}} = 3.7\times10^6$ (Fig. 3(f)) [47].

Next, we consider an *out-of-plane* AFTJ, where electron transport occurs perpendicular to the plane of the junction composed of the $IrTe_2/In_2S_3/PtTe_2$ heterostructure (Fig. 4(a)). Here, $IrTe_2$ and $PtTe_2$ serve as electrodes, which are known to be metallic van der Waals materials and have a small lattice mismatch with $In_2S_3$ [71,72]. In this AFTJ, $IrTe_2$ and $PtTe_2$ are expected to provide boundary conditions for polarization charge screening in $In_2X_3$ and hence to strongly affect the transmission across the AFTJ for different polarization states. Figs. 4(b-e) show the calculated LDOS of the $IrTe_2/In_2S_3/PtTe_2$ heterostructure. It is seen that for the AFE-T state, the $E_F$ is located at a lower energy within the band gap of $In_2S_3$ compared to that for the other polarization states. This is due to the bound charge at the center of the $In_2S_3$ bilayer, which shifts the potential energy of the $In_2S_3$ bands higher with respect to $E_F$ determined by the electrodes. Therefore, the AFE-T state has the lowest transmission at $E_F$ (Fig. 4(f)). On the contrary, for the AFE-H state, the $E_F$ is located inside the CBM for both $In_2S_3$ layers (Fig. 4(c)), leading to the highest transmission among the four states (Fig. 4(f)). For the FE states, although there is screening of the polarization charges by the top and bottom metal electrodes, the depolarizing electric field and the relative band shift of the two layers cannot be fully eliminated. As seen from Figs. 4(d-e) and Fig. S7 [47], for both FE states, the $E_F$ is located in the gap of the top $In_2S_3$ layer, and in the conduction band of the bottom $In_2S_3$ layer. The transmission of the AFTJ for the FE-down state is larger than that for the FE-up state, due to the $E_F$ being higher in energy with respect to the CBM of the bottom $In_2S_3$ layer. We find the on/off transmission ratios for the out-of-plane AFTJ are smaller than those for the in-plane AFTJ. This can be understood from the barrier width being nearly five times smaller for the out-of-plane AFTJ (~1.5 nm) compared to the in-plane AFTJ (~7 nm) considered in this work. Inserting suitable insulating buffer layers between the electrodes and the bilayer $In_2S_3$ can increase the barrier width and prevent the hybridization of the $In_2S_3$ and electrodes, which is expected to enhance the on/off ratios for four non-volatile resistance states.

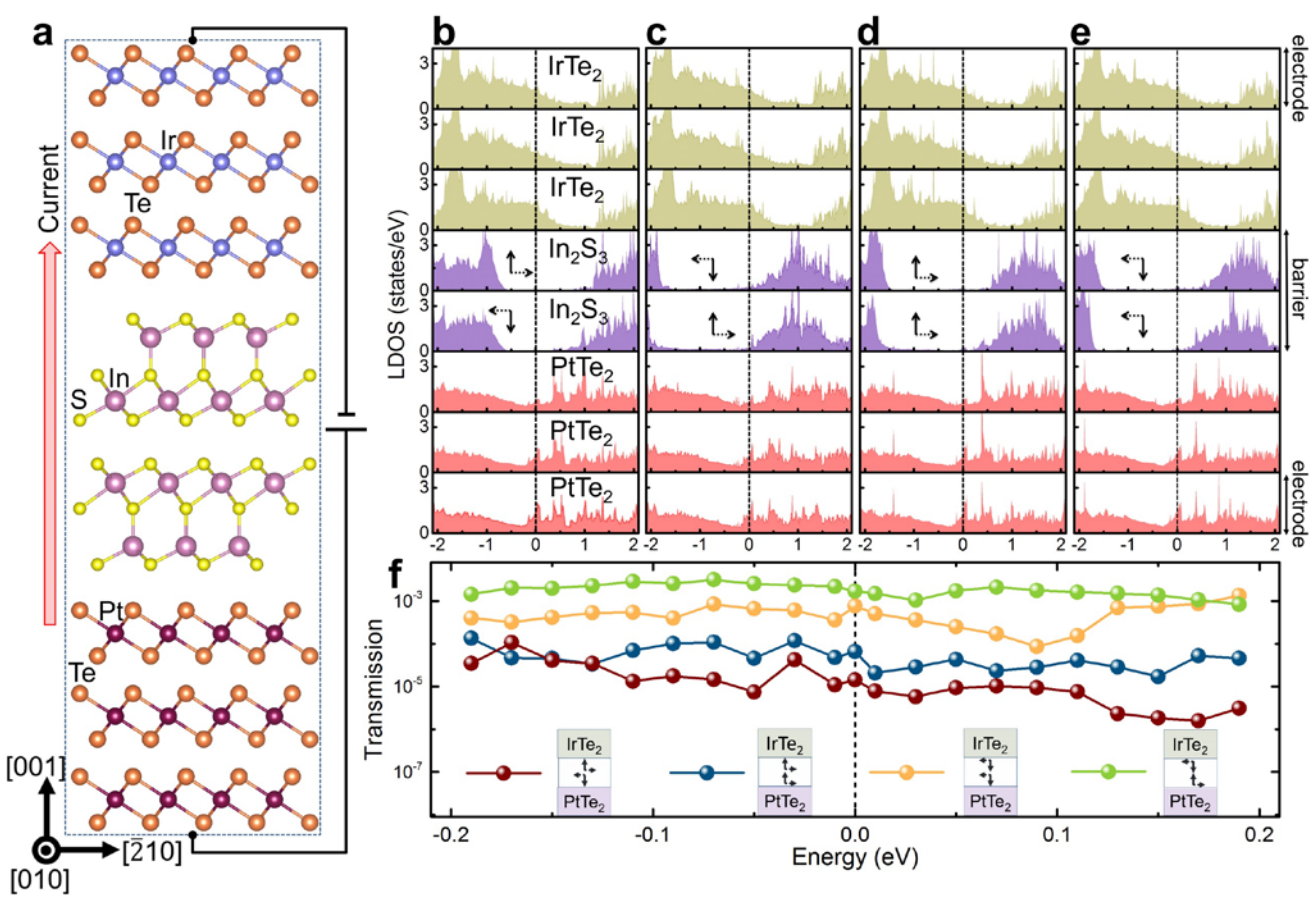

**FIG. 4. (a)** Atomic structure of the $IrTe_2/In_2S_3/PtTe_2$ out-of-plane AFTJ in the AFE-T state. (**b-e**) The layer-resolved density of states (LDOS) of the $IrTe_2/In_2S_3/PtTe_2$ out-of-plane AFTJ in the AFE-T state (**b**), AFE-H state (**c**), and FE state with positive (**d**) and negative (**e**) $P_\perp$. (**f**) Transmission as a function of electron energy for the $IrTe_2/In_2S_3/PtTe_2$ out-of-plane AFTJ.

We find a large contrast between the multiple resistance states being preserved in the presence of finite bias (Fig. S11) [47]. Similarly, the multiple resistance states are expected in the AFTJs based on FE $In_2Se_3$ and $In_2Te_3$ barriers. Switching between different polarization states can be realized by applying suitable out-of-plane electric field [47]. Recent discoveries of new van der Waals materials and progress in fabrication of van der Waals heterostructures allows the realization of in-plane and out-of-plane AFTJs with engineered interfaces and designed functionalities.

In conclusion, we have demonstrated new functionalities offered by 2D FE van der Waals materials if they are exploited as tunnel barriers. Due to weak coupling between the monolayers in these materials, the relative dipole alignment between them can be controlled by applied voltage. This allows transitions between FE and AFE states, resulting in the change of the barrier height and thus transmission across the tunnel junction. We have explored these functionalities by considering 2D AFTJs based on bilayer $In_2X_3$ (X = S, Se, Te) barriers and different electrodes, and predicted the appearance of giant TER effects and multiple non-volatile resistance states driven by AFE-FE order transitions. Our proposal opens a new route to realize the nanoscale memory devices with ultrahigh storage density using 2D AFTJs.

**Acknowledgments.** This work was supported by the National Science Foundation (NSF) through the Nebraska MRSEC (Grant DMR-1420645). J.D. was supported by the China Scholarship Council and the National Natural Science Foundation of China (Grant Nos. 11604078 and 11347187). Computations were performed at the University of Nebraska Holland Computing Center.

* These authors contributed equally to this work.
† dfshao@unl.edu
‡ tsymbal@unl.edu

[1] J. Frenkel, On the electrical resistance of contacts between solid conductors. *Phys. Rev.* **36**, 1604–1618 (1930).
[2] M. Y. Zhuravlev, R. F. Sabirianov, S. S. Jaswal and E. Y. Tsymbal, Giant electroresistance in ferroelectric tunnel junctions. *Phys. Rev. Lett.* **94**, 246802 (2005).
[3] H. Kohlstedt, N. Pertsev, J. R. Contreras and R. Waser, Theoretical current-voltage characteristics of ferroelectric tunnel junctions. *Phys. Rev. B* **72**, 125341 (2005).
[4] E. Y. Tsymbal and H. Kohlstedt, Tunneling across a ferroelectric. *Science* **313**, 181–183 (2006).
[5] A. Chanthbouala, A. Crassous, V. Garcia, K. Bouzehouane, S. Fusil, X. Moya, J. Allibe, B. Dlubak, J. Grollier, S. Xavier and C. Deranlot, Solid-state memories based on ferroelectric tunnel junctions. *Nat. Nanotechnol.* **7**, 101-104 (2012).
[6] V. Garcia and M. Bibes, Ferroelectric tunnel junctions for information storage and processing. *Nat. Commun.* **5**, 4289 (2014).
[7] J. P. Velev, J. D. Burton, M. Y. Zhuravlev and E. Y. Tsymbal, Predictive modelling of ferroelectric tunnel junctions. npj Comput. Mater. **2**, 16009 (2016).
[8] Z. Wen and D. Wu, Ferroelectric tunnel junctions: Modulations on the potential barrier. *Adv. Mater.* 1904123 (2019); DOI: 10.1002/adma.201904123
[9] D. Pantel, H. Lu, S. Goetze, P. Werner, D. J. Kim, A. Gruverman, D. Hesse, and M. Alexe, Tunnel electroresistance in junctions with ultrathin ferroelectric $Pb(Zr_{0.2}Ti_{0.8})O_3$ barriers. *Appl. Phys. Lett.* **100**, 232902 (2012).
[10] D. J. Kim, H. Lu, S. Ryu, C. W. Bark, C. B. Eom, E. Y. Tsymbal, and A. Gruverman, Ferroelectric tunnel memristor. *Nano Lett.* **12**, 5697-5702 (2012).
[11] Z. Wen, C. Li, D. Wu, A. Li, and N. Ming, Ferroelectric-field-effect-enhanced electroresistance in metal/ferroelectric/semiconductor tunnel junctions. *Nat. Mater.* **12**, 617 (2013).
[12] E. Y. Tsymbal and A. Gruverman, Beyond the barrier. *Nat. Mater.* **12**, 602 (2013).
[13] R. Soni, A. Petraru, P. Meuffels, O. Vavra, M. Ziegler, S.K. Kim, D.S. Jeong, N. A. Pertsev, and H. Kohlstedt, Giant electrode effect on tunnelling electroresistance in ferroelectric tunnel junctions. *Nat. Commun.* **5**, 5414 (2014).
[14] G. Radaelli, D. Gutiérrez, F. Sánchez, R. Bertacco, M. Stengel, and J. Fontcuberta, Large room-temperature electroresistance in dual-modulated ferroelectric tunnel barriers. *Adv. Mater.* **27**, 2602-2607 (2015).
[15] X. Liu, J. D. Burton and E. Y. Tsymbal, Enhanced tunneling electroresistance in ferroelectric tunnel junctions due to the reversible metallization of the barrier. *Phys. Rev. Lett.* **116**, 197602 (2016)
[16] T. Li, P. Sharma, A. Lipatov, H. Lee, J. W. Lee, M.Y. Zhuravlev, T.R. Paudel, Y.A. Genenko, C.B. Eom, E.Y. Tsymbal and A. Sinitskii, Polarization-mediated modulation of electronic and transport properties of hybrid $MoS_2$–$BaTiO_3$–$SrRuO_3$ tunnel junctions. *Nano Lett.* **17**, 922-927 (2017).
[17] Z. Xi, J. Ruan, C. Li, C. Zheng, Z. Wen, J. Dai, A. Li and D. Wu, Giant tunnelling electroresistance in metal/ferroelectric/semiconductor tunnel junctions by engineering the Schottky barrier. *Nat. Commun.* **8**, 15217 (2017).
[18] A. Tsurumaki-Fukuchi, H. Yamada and A. Sawa, Resistive switching artificially induced in a dielectric/ferroelectric composite diode. *Appl. Phys. Lett.* **103**, 152903 (2013).
[19] H. Lu, A. Lipatov, S. Ryu, D. J. Kim, H. Lee, M. Y. Zhuravlev, C.B. Eom, E.Y. Tsymbal, A. Sinitskii, and A. Gruverman, Ferroelectric tunnel junctions with graphene electrodes. *Nat. Commun.* **5**, 5518 (2014).
[20] V. S. Borisov, S. Ostanin, S. Achilles, J. Henk and I. Mertig, Spin-dependent transport in a multiferroic tunnel junction: Theory for $Co/PbTiO_3/Co$. *Phys. Rev. B* **92**, 075137 (2015).
[21] C. Li, L. Huang, T. Li, W. Lü, X. Qiu, Z. Huang, Z. Liu, S. Zeng, R. Guo, Y. Zhao and K. Zeng, Ultrathin $BaTiO_3$-based ferroelectric tunnel junctions through interface engineering. *Nano Lett.* **15**, 2568-2573 (2015).
[22] S. A. Fedorov, A. E. Korolkov, N. M. Chtchelkatchev, O. G. Udalov, and I. S. Beloborodov, Interplay of ferroelectricity and single electron tunneling. *Phys. Rev. B* **89**, 155410 (2014).
[23] G. Sanchez-Santolino, J. Tornos, D. Hernandez-Martin, J. I. Beltran, C. Munuera, M. Cabero, A. Perez-Muñoz, J. Ricote, F. Mompean, M. Garcia-Hernandez, Z. Sefrioui, C. Leon, S. J. Pennycook, M. C. Muñoz, M. Varela, and J. Santamaria, Resonant electron tunneling assisted by charged domain walls in multiferroic tunnel junctions. *Nat. Nanotechnol.* **12**, 655-662 (2017).
[24] M. Li, L. L. Tao, J. P. Velev, and E. Y. Tsymbal, Resonant tunneling across a ferroelectric domain wall. *Phys. Rev. B* **97**, 155121 (2018).
[25] M. Li, L. L. Tao, and E. Y. Tsymbal, Domain-wall tunneling electroresistance effect. *Phys. Rev. Lett.* **123**, 266602 (2019).
[26] C. Kittel, Theory of antiferroelectric crystals. *Phys. Rev.* **82**, 729-732 (1951).
[27] X. Tan, C. Ma, J. Frederick, S. Beckman and K. G. Webber, The antiferroelectric-ferroelectric phase transition in lead-containing and lead-free perovskite ceramics. *J. Am. Ceram. Soc.* **94**, 4091-4107 (2011).
[28] W. Pan, Q. Zhang, A. Bhalla and L. E. Cross, Field-forced antiferroelectric-to-ferroelectric switching in modified lead

zirconate titanate stannate ceramics. *J. Am. Ceram. Soc.* **72**, 571-578 (1989).

[29] G. Apachitei, J. J. P. Peters, A. M. Sanchez, D. J. Kim and M. Alexe, Antiferroelectric tunnel junctions. *Adv. Electron. Mater.* **3**, 1700126 (2017)

[30] W. Ding, J. Zhu, Z. Wang, Y. Gao, D. Xiao, Y. Gu, Z. Zhang and W. Zhu, Prediction of intrinsic two-dimensional ferroelectrics in $In_2Se_3$ and other $III_2$-$VI_3$ van der Waals materials. *Nat. Commun.* **8**, 14956 (2017).

[31] C. Cui, F. Xue, W.-J. Hu and L.-J. Li, Two-dimensional materials with piezoelectric and ferroelectric functionalities. *npj 2D Mater. Appl.* **2**, 18 (2018).

[32] A. Belianinov, Q. He, A. Dziaugys, P. Maksymovych, E. Eliseev, A. Borisevich, A. Morozovska, J. Banys, Y. Vysochanskii, and S.V. Kalinin, $CuInP_2S_6$ room temperature layered ferroelectric. *Nano Lett.* **15**, 3808-3814 (2015).

[33] F. Liu, L. You, K. L. Seyler, X. Li, P. Yu, J. Lin, X. Wang, J. Zhou, H. Wang, H. He, and S. T. Pantelides, Room-temperature ferroelectricity in $CuInP_2S_6$ ultrathin flakes. *Nat. Commun.* **7**, 12357 (2016).

[34] K. Chang, J. Liu, H. Lin, N. Wang, K. Zhao, A. Zhang, F. Jin, Y. Zhong, X. Hu, W. Duan and Q. Zhang, Discovery of robust in-plane ferroelectricity in atomic-thick SnTe. *Science* **353**, 274-278 (2016).

[35] K. Liu, J. Lu, S. Picozzi, L. Bellaiche, and H. Xiang, Intrinsic origin of enhancement of ferroelectricity in SnTe ultrathin films. *Phys. Rev. Lett.* **121**, 027601 (2018).

[36] K. Chang, T.P. Kaloni, H. Lin, A. Bedoya-Pinto, A.K. Pandeya, I. Kostanovskiy, K. Zhao, Y. Zhong, X. Hu, Q. K. Xue, and X. Chen, Enhanced spontaneous polarization in ultrathin SnTe films with layered antipolar structure. *Adv. Mater.* **31**, 1804428 (2019).

[37] J. Xiao, H. Zhu, Y. Wang, W. Feng, Y. Hu, A. Dasgupta, Y. Han, Y. Wang, D. A. Muller, L. W. Martin, P. A. Hu, and X. Zhang, ntrinsic two-dimensional ferroelectricity with dipole locking. *Phys. Rev. Lett.* **120**, 227601 (2018).

[38] C. Cui, W.-J. Hu, X. Yan, C. Addiego, W. Gao, Y. Wang, Z. Wang, L. Li, Y. Cheng, P. Li, X. Zhang, H. N. Alshareef, T. Wu, W. Zhu, X. Pan, and L.-J. Li, Intercorrelated in-plane and out-of-plane ferroelectricity in ultrathin two-dimensional layered semiconductor $In_2Se_3$. *Nano Lett.* **18**, 1253-1258 (2018).

[39] Y. Zhou, D. Wu, Y. Zhu, Y. Cho, Q. He, X. Yang, K. Herrera, Z. Chu, Y. Han, M. C. Downer, H. Peng, and K. Lai, Out-of-plane piezoelectricity and ferroelectricity in layered α-$In_2Se_3$ nanoflakes. *Nano Lett.* **17**, 5508-5513 (2017).

[40] F. Xue, W. Hu, K. C. Lee, L.-S. Lu, J. Zhang, H.-L. Tang, A. Han, W. T. Hsu, S. Tu, W.-H. Chang, C.-H. Lien, J.-H. He, Z. Zhang, L.-J. Li, and X. Zhang, Room-temperature ferroelectricity in hexagonally layered α-$In_2Se_3$ nano flakes down to the monolayer limit. *Adv. Funct. Mater.* **28**, 1803738 (2018).

[41] S. Wan, Y. Li, W. Li, X. Mao, W.Zhu, and H. Zeng, Room-temperature ferroelectricity and a switchable diode effect in two-dimensional α-$In_2Se_3$ thin layers. *Nanoscale* **10**, 14885-14892 (2018).

[42] F. Xue, J. Zhang, W. Hu, W. T. Hsu, A., Han, S. F. Leung, J. K. Huang, Y. Wan, S. Liu, J. Zhang, and J. H. He, Multidirection piezoelectricity in mono and multilayered hexagonal α-$In_2Se_3$. *ACS Nano* **12**, 4976-4983 (2018).

[43] M. Dai, Z. Wang, F. Wang, Y. Qiu, J. Zhang, C.Y. Xu, T. Zhai, W. Cao, Y. Fu, D. Jia, and Y. Zhou, Two-dimensional van der Waals materials with aligned in-plane polarization and large piezoelectric effect for self-powered piezoelectric sensors. *Nano Lett.* **19**, 5410-5416 (2019).

[44] The antiferroelectric state here refers to the state with the polarization of each van der Waals monolayer is antiparallel to that of the adjacent monolayer. This state has been previously considered to be unstable in three-dimensional perovskite ferroelectrics [45,46].

[45] W. Kinase, On interactions among ions of a $BaTiO_3$ Crystal and on its 180° and 90° type domain boundaries. *Prog. Theor. Phys.***13**, 529 (1955).

[46] W. Kinase, K.Yano, and N. Ohnishi. Dipole interactions in antiferroelectric $PbZrO_3$. *Ferroelectrics* **46**, 281 (1983).

[47] See Supplemental Material for calculation methods, atomic structures and calculated properties of bilayer $In_2X_3$ (X = S, Se, Te), calculated electronic and transport properties of $(Cd_{0.5}In_{0.5})_2S_3/In_2S_3/(Sn_{0.5}In_{0.5})_2S_3$ in-plane AFTJ, band structure of $IrTe_2/In_2S_3/PtTe_2$ heterostructure, discussions of the electric switching of different states in bilayer $In_2X_3$ (X = S, Se, Te), and the non-equilibrium transport properties of the $In_2S_3$ based AFTJs, which includes Refs. [48–61].

[48] G. Kresse and D. Joubert, From ultrasoft pseudopotentials to the projector augmented-wave method. *Phys. Rev. B* **59**, 1758-1775 (1999).

[49] G. Kresse and J. Furthmuller, Efficient iterative schemes for ab initio total-energy calculations using a plane-wave basis set. *Phys. Rev. B* **54**, 11169-11186 (1996).

[50] S. Grimme, J. Antony, S. Ehrlich, and H. Krieg, A consistent and accurate ab initio parametrization of density functional dispersion correction (DFT-D) for the 94 elements H-Pu. *J. Chem. Phys.* **132**, 154104 (2010).

[51] L. Bengtsson, Dipole correction for surface supercell calculations. *Phys. Rev. B* **59**, 12301-12304 (1999).

[52] G. Henkelman, B. P. Uberuaga, and H. Jonsson, A climbing image nudged elastic band method for finding saddle points and minimum energy paths. *J. Chem. Phys.* **113**, 9901-9904 (2000).

[53] J. Taylor, H. Guo, J. Wang, Ab initio modeling of quantum transport properties of molecular electronic devices. *Phys. Rev. B* **63**, 245407 (2001).

[54] M. Brandbyge, J.-L. Mozos, P. Ordejón, J. Taylor, and K. Stokbro, Density-functional method for nonequilibrium electron transport. *Phys. Rev. B* **65**, 165401 (2002).

[55] ATOMISTIX TOOLKIT version 2015.1 Synopsys QuantumWise (www.quantumwise.com). QuantumWise A/S is now part of Synopsys, and from the upcoming version ATK will be part of the QuantumATK suite.

[56] T. Ozaki, Variationally optimized atomic orbitals for large-scale electronic structures. *Phys. Rev. B* **67**, 155108 (2003).

[57] T. Ozaki and H. Kino, Numerical atomic basis orbitals from H to Kr. *Phys. Rev. B* **69**, 195113 (2004).

[58] G. Henkelman, A. Arnaldsson, and H. Jónsson, A fast and robust algorithm for Bader decomposition of charge density. *Comput. Mater. Sci.* **36**, 354-360 (2006).

[59] W. Tang, E. Sanville, and G. Henkelman, A grid-based Bader analysis algorithm without lattice bias. *J. Phys.: Condens. Matter* **21**, 084204 (2009).

[60] M.-J. Spijkman, K. Myny, E. C. P. Smits, P. Heremans, P. W. M. Blom, and D. M. de Leeuw, Dual-gate thin-film transistors, integrated circuits and sensors, *Adv. Mater.* **23**, 3231 (2011).

[61] S. Datta, Electronic transport in mesoscopic systems (Cambridge University Press, 1997).

[62] S. Popović, B. Čelustka, and D. Bidjin, X-ray diffraction measurement of lattice parameters of $In_2Se_3$. *Phys. Stat. Sol. (a)* **6**, 301-304 (1971).

[63] S. Popović, A. Tonejc, B. Čelustka, B. Grzeta-Plenkovic, and R. Trojko, Revised and new crystal data for indium selenides. *J. Appl. Cryst.* **12**, 416-420 (1979).

[64] Y. Zhang, J. Sun, J. P. Perdew, and X. Wu, Comparative first-principles studies of prototypical ferroelectric materials by LDA, GGA, and SCAN meta-GGA. *Phys. Rev. B* **96**, 035143 (2017).

[65] M. Li, L. L. Tao, and E. Y. Tsymbal, Domain-wall tunneling electroresistance effect, *Phys. Rev. Lett.* **123**, 266602 (2019).

[66] J. P. Perdew, K. Burke, and M. Ernzerhof, Generalized gradient approximation made simple. *Phys. Rev. Lett.* **77**, 3865-3868 (1996).

[67] A. V. Krukau, O. A. Vydrov, A. F. Izmaylov, and G. E. Scuseria, Influence of the exchange screening parameter on the performance of screened hybrid functionals. *J. Chem. Phys.* **125**, 224106 (2006).

[68] X. Shen, Y.-W. Fang, B. Tian, and C.-G. Duan, Two-dimensional ferroelectric tunnel junction: The case of Monolayer In:SnSe/SnSe/Sb:SnSe homostructure. *ACS Appl. Electron. Mater.* **1**, 1133 (2019).

[69] L. Kang, P. Jiang, N. Cao, H. Hao, X. Zheng, L. Zhang, and Z. Zeng, Realizing giant tunneling electroresistance in two-dimensional graphene/BiP ferroelectric tunnel junction. *Nanoscale* **11**, 16837 (2019).

[70] L. Kang, P. Jiang, H. Hao, Y. Zhou, X. Zheng, L. Zhang, and Z. Zeng, Giant tunneling electroresistance in two-dimensional ferroelectric tunnel junctions with out-of-plane ferroelectric polarization. *Phys. Rev. B* **101**, 014105 (2020).

[71] N. Matsumoto, K. Taniguchi, R. Endoh, H. Takano, and S. Nagata, Resistance and susceptibility anomalies in $IrTe_2$ and $CuIr_2Te_4$. *J. Low. Temp. Phys.* **117**, 1129-1133 (1999).

[72] P. J. Orders, J. Lieseǵanǵ, R. C. G. Leckey, J. G. Jenkin, and J. D. Riley Angle-resolved photoemission from the valence bands of $NiTe_2$, $PdTe_2$ and $PtTe_2$. *J. Phys. F: Met. Phys.* **12**, 2737 (1982).